\documentclass{aa}  

\usepackage{graphicx}
\usepackage{txfonts}
\usepackage{gensymb}
\usepackage{threeparttable}
\usepackage{xcolor}
\usepackage{lastpage}
\usepackage{scalerel}
\newcommand{\kms}[0]{km s$^{-1}$}

\usepackage{hyperref}
\hypersetup{colorlinks=true,
    citecolor=blue,
    linkcolor=blue,
    urlcolor=blue,
    }

\usepackage{cleveref}

\begin{document}

   \title{SN 2024ggi: another year, another striking Type II supernova}

   \author{K. Ertini \inst{1,2} 
          \and T. A. Regna \inst{1,2} 
          \and L. Ferrari \inst{1,2} \and M. C. Bersten \inst{1,2,3} \and G. Folatelli \inst{1,2,3} \and A. Mendez Llorca \inst{1,2} \and E. Fernández-Lajús \inst{1} \and G. A. Ferrero \inst{2,1} 
          \and E. Hueichapán Díaz \inst{4,5} \and R. Cartier \inst{4} \and L. M. Román Aguilar \inst{1,2} \and C. Putkuri \inst{1} \and M. P. Piccirilli \inst{2,6} \and S. A. Cellone \inst{7,2} \and J. Moreno \inst{1,2} \and M. Orellana \inst{8,6}
          \and J. L. Prieto \inst{4,5} \and M. Gerlach \inst{4,5,9} \and V. Acosta \inst{2} \and M. J. Ritacco \inst{2} \and J. C. Schujman \inst{2} \and J. Valdéz \inst{2}
          }

   \institute{Instituto de Astrofísica de La Plata (IALP), CCT-CONICET-UNLP, Paseo del Bosque, B1900FWA La Plata, Argentina\\
              \email{keilaertini@fcaglp.unlp.edu.ar}
        \and
        Facultad de Ciencias Astronómicas y Geofísicas, Universidad Nacional de La Plata, Paseo del Bosque S/N, 1900 La Plata, Buenos Aires, Argentina
        \and
        Kavli Institute for the Physics and Mathematics of the Universe (WPI), The University of Tokyo, 5-1-5 Kashiwanoha, Kashiwa, Chiba 277-8583, Japan 
        \and
        Instituto de Estudios Astrofísicos, Facultad de Ingeniería y Ciencias, Universidad Diego Portales, Av. Ejército Libertador 441, Santiago, Chile
        \and
        Millennium Institute of Astrophysics MAS, Nuncio Monseñor Sotero Sanz 100, Off. 104, Providencia, Santiago, Chile 
        \and
        Consejo Nacional de Investigaciones Científicas y Técnicas (CONICET), Godoy Cruz 2290, 1425 Ciudad Autónoma de Buenos Aires, Argentina 
        \and
        Complejo Astronómico El Leoncito (CASLEO), CONICET - UNLP - UNC - UNSJ, Av. España 1512 (sur), J5402DSP San Juan, Argentina 
        \and
        Universidad Nacional de Río Negro. Sede Andina, Laboratorio de Investigación Científica en Astronomía, Anasagasti 1463, Bariloche (8400), Argentina 
        \and
        Instituto de Astrof\'isica, Facultad de F\'isica, Pontificia Universidad Cat\'olica de Chile, Av. Vicuña Mackenna 4860, Santiago, Chile 
             }

   \date{Received Month XX, XXXX; accepted Month XX, XXXX}

 
   \abstract
   {SN~2024ggi is a Type II supernova that exploded in the nearby galaxy NGC 3621 at a distance of approximately 7 Mpc, making it one of the closest supernovae of the decade. This SN shows clear signs of interaction with a dense circumstellar material, and several studies have investigated the properties of its possible progenitor star using pre-explosion data.
   }
   {In this work we aim to constrain the progenitor properties of SN~2024ggi by performing hydrodynamical modeling of its bolometric light curve and expansion velocities using our own spectrophotometric data.
   }
   {We present photometric and spectroscopic observations of SN~2024ggi obtained in the Complejo Astronómico El Leoncito, in Las Campanas Observatory, and in Las Cumbres Observatory Global Telescope Network, spanning from 2 to 106 days after explosion. We constructed its bolometric light curve and we characterize it by calculating its morphological parameters. 
   Then, we computed a grid of one dimensional explosion models for evolved stars with varying masses and estimated the properties of the progenitor star of SN~2024ggi by comparing the models to the observations.
   }
   {The observed bolometric luminosity and expansion velocities are well-matched by a model involving the explosion of a star in the presence of a close circumstellar material (CSM), with a zero-age main sequence mass of $\mathrm{M_{ZAMS}}$ = 15 $M_{\odot}$, a pre-SN mass and radius of 14.1 $M_{\odot}$ and 517 $R_{\odot}$, respectively, an explosion energy of $1.3\times10^{51}$ erg, and a nickel mass below 0.035 $M_{\odot}$. Our analysis suggests that the progenitor suffered a mass-loss rate of $4 \times 10^{-3}$ $M_{\odot}$~yr$^{-1}$, confined to a distance of 3000 $R_{\odot}$. The CSM distribution is likely a two-component structure that consists of a compact core and an extended tail. This analysis represents the first hydrodynamical model of SN~2024ggi with a complete coverage of the plateau phase. 
   }
   {}


   \maketitle

\section{Introduction}
\label{sec:intro}
Type II supernovae (SNeII) mark the end of the life of massive stars ($\gtrapprox~8$ $M_{\odot}$) that have retained their hydrogen envelopes. A subclass of SNeII are Type IIn SNe, which show narrow lines in their spectra, coming from the ejecta colliding with circumstellar material (CSM: \citealt{schlegel90}).  In the last decade, observations of some SNeII revealed early spectra showing narrow lines of highly ionized material, typical of IIn SNe, which disappear hours to days later, thus becoming normal SNeII afterward \citep{yaron17}. With modern high-cadence surveys discovering younger SNe, just a few hours after their first light, this situation became increasingly usual. Recently, \citet{bruch23} analyzed a sample of 40 SNeII with good early spectral coverage (spectra obtained within $\lesssim$ 2 days from the explosion epoch) and found that 40 \% of them showed flash ionized features, suggesting that this is a rather common phenomenon. Modelling of the SNe showing flash features suggests that their RSG progenitors ejected material at an unexpectedly high mass loss rate prior to the explosion \citep{yaron17,dessart16}. 

On 2023, the Type II SN~2023ixf \citep{perley23} was discovered in the M101 galaxy \citep{itagaki23}. Its proximity and early detection made it possible to obtain a vast amount of observations across multiple wavelengths. Particularly, flash spectra were taken less than a day after discovery showing narrow features of H I, He I/II, C IV, and N III/IV/V, as a sign of interaction with a dense CSM \citep{jacobsongalan23}. In addition, SN~2023ixf was an excellent opportunity to study the environment and progenitor system of type II SNe in detail. Pre-explosion observations of the SN~2023ixf site suggest a dusty red supergiant progenitor, with zero-age main sequence (ZAMS) mass estimates ranging from 8 to 18 $M_{\odot}$ \citep{qin23,kilpatrick23,jencson23,xiang24,neustadt24,vandyk24}. Including additional methods, such as variability studies of the pre-SN source, stellar population analyses, hydrodynamical modeling, and nebular spectroscopy, expands the estimated ZAMS mass range to 8–22 $M_{\odot}$ \citep{soraisam23,niu23,liu23,bersten24,ferrari24}.

In 2024, another very nearby Type II SN, SN~2024ggi, was discovered by the Asteroid Terrestrial-impact Last Alert System (ATLAS; \citealp[]{tonry18}) on 11 Apr, 2024 (MJD = 60411.14) with a magnitude of 18.92 $\pm$ 0.08 mag in the ATLAS $o$-band \citep{srivastav24}. The host galaxy of SN~2024ggi is NGC 3621 at a distance of 6.7 Mpc \citep{paturel02}. It was spectroscopically classified by \citet{zhai24}, and its last non detection published by \citet{killestein24} was on MJD = 60410.45, i.e. only 0.69 days before the detection. A spectrum taken the day of the discovery revealed flash ionized features \citep{hoogendam24}. Since then, comprehensive ultraviolet and optical photometry, as well as spectroscopy was obtained with high cadence \citep{jacobsongalan24,thallis24,chen24,shrestha24}. Follow up observations across other wavelengths were also triggered, including detections in X-rays \citep{zhangJ24,margutti24} and radio \citep{ryder24}, as well as non-detections the centimeter and millimeter regimes, \citep{chandra24,hu24}, and $\gamma$-rays\citep{marti-devesa24}. Flash spectra were studied in detail by \citet{jacobsongalan24}, \citet{thallis24}, and \citet{shrestha24}. Just hours after discovery, emission lines of the Balmer series, He~I, C~III, and N~III were detected, and less than a day after that emission lines of He~II, C~IV, N IV/V, and O~V became visible. This rise on ionization was accompanied by an evolution toward bluer colors. The duration of the flash features was $3.8 \pm 1.6$ days \citep{jacobsongalan24}.

Given its proximity, SN~2024ggi offers yet another valuable opportunity to probe the progenitors of Type II SNe. Shortly after its discovery, several telegrams reported searches of its progenitor in archival data. \citet{komura24} examined XMM-Newton archival observations for X-ray emission but identified no apparent X-ray source at the SN position. Additionally,  \citet{srivastav24} and \citet{yang24} reported a possible red progenitor source in archival data from the Legacy Survey and Hubble Space Telescope (HST), and \citet{perez-fournon24} identified a likely progenitor in near-IR archival imaging and catalogs of the VISTA Hemisphere Survey. 

Following these telegrams, further studies analyzed archival data to investigate the progenitor. Using pre-explosion images from the HST and the \textit{Spitzer} Space Telescope, \citet{xiang24} inferred that the progenitor of SN~2024ggi was a red bright variable star with a pulsational period of approximately 379 days in the mid-infrared. 
 By fitting the progenitor's spectral energy distribution, they derived an initial mass of $\mathrm{13~M_{\odot}}$. Additionally, \citet{chen24} used archival deep images from the Dark Energy Camera Legacy Survey, and suggested a possible progenitor with an initial mass in the range of $\mathrm{14-17~M_{\odot}}$. 

Independently of direct progenitor detections, \citet{hong24} performed an environmental analysis based on images from the HST. They determined that the age of the youngest stellar population in the environment of the SN, associated with the progenitor, permits to derive for the latter an initial mass of $\mathrm{10.2~M_{\odot}}$.

In this paper, we adopt an alternative approach to estimate the progenitor mass of SN~2024ggi by comparing hydrodynamical explosion models with the bolometric light curve and expansion velocity evolution of the SN. This work represents the first attempt to do so using data during the full extent of the plateau phase. Additionally, we present photometric and spectroscopic follow-up of SN~2024ggi starting 5 days after the explosion. The paper is organized as follows. In Section \ref{sec:data} we present the observations and data reduction of SN~2024ggi. We analyse its photometric and spectroscopic properties, as well as its bolometric light curve in Section \ref{sec:prop}, and in Section \ref{sec:hydro} we present the associated hydrodynamical modelling for the proposed progenitor scenario. Finally, in Section \ref{sec:conclusions} we provide a summary of our results.

\section{Data sample}
\label{sec:data}

We obtained direct images with the 60 cm Helen Sawyer Hogg (HSH) telescope at the Complejo Astronómico El Leoncito (CASLEO), located in San Juan, Argentina, through the programs HSH-2024A-DD01 (PI Ertini) and HSH-2024A-02 (PI Fernández-Lajús). The observations were obtained with a nearly daily cadence, from 5 to 35 days post explosion. We used the $B$, $V$, $R$ and $I$ filters. The observations were divided into 10 exposures of 40 seconds each for $VRI$ bands and 60 seconds each for the $B$ band. Sky flats were taken each day of observation. The reduction was performed following standard procedures in Python. The photometry was performed using the software AutoPhOT \citep{Brennan2022}. The software determines whether to utilize aperture or point-spread function (PSF) photometry. It starts with aperture photometry as an initial estimate to determine the approximate magnitude of the SN, then attempts to apply PSF photometry. By default, the source must have a signal-to-noise ratio greater than 25 to be considered for the PSF model. If PSF fitting is not feasible, aperture photometry is used instead. The instrumental magnitudes are then calibrated to the standard system estimating a zero point, which is calculated using $\sim$20 Pan-STARRS field stars \citep{flewelling20}. Since the Pan-STARRS magnitudes are in $gri$ we first used the transformation coefficients of \citet{tonry18} to transform the magnitudes of the field stars into $BVRI$.  The photometry is listed in Table \ref{tab:phot}. 

We also obtained images using the 1~m telescope of Las Cumbres Observatory Global Telescope Network (LCOGTN) located at Cerro Tololo. These observations were conducted from 2 to 85 days post-explosion, with a typical cadence of one observation approximately every three days. There was 12-day gap in coverage due to adverse weather conditions. Additionally, a final observation was taken 106 days after the explosion. The reduction of LCOGT data was done by using a new python version of the custom IRAF \citep{tody86} script package described in \citet{hamuy06} and \citet{contreras10}. LCOGT photometry is listed in Table \ref{tab:phot_griz}.

Spectra were acquired with the REOSC spectrograph mounted on the 2.15 m Jorge Sahade (JS) telescope (program JS-2024A-DD01, PI Ertini) at CASLEO. We used the 200-mic slit and the \#270 grating, covering a wavelength range of 3400-7600 \AA. The data reduction included wavelength and flux calibration using arc lamps and spectrophotometric standard stars, respectively. The spectral resolution measured from skylines in the spectra is $\approx 8$ \AA, which results in $\approx 480$ km s$^{-1}$ at 5000 \AA. We also obtained one spectrum using the LDSS-3 spectrograph mounted at the 6.5~m Magellan Clay telescope at Las Campanas Observatory on July 3, 2024 at 23:18:40 UT (MJD = 60494.97). The SN was observed at the parallactic angle using the 1´´ slit width in combination with the VPH-ALL 400 lines/mm grism providing a resolution of ~8.2 \AA~(R=900). A HeArNe comparison lamp was obtained immediately after the SN spectrum to perform the wavelength calibration. The spectrum was reduced and flux calibrated using IRAF routines (see \citealp{cartier24}). The log of spectroscopic observations is listed in Table \ref{tab:log_spec}.

\begin{table}
	\caption{Log of spectroscopic observations. The phase is indicated in rest-frame days from explosion.}
	\label{tab:log_spec}
	\begin{tabular}{lcccc} 
		\hline
		Date & MJD & Telescope & Instrument & Phase  \\
		\hline
		2024 Apr 16 & 60416.05 & JS & REOSC & 5.3 \\ 
		2024 Apr 23 & 60423.98 & JS & REOSC & 13.2  \\
		2024 Apr 24 & 60424.97 & JS & REOSC & 14.2  \\
		2024 Apr 25 & 60425.98 & JS & REOSC & 15.2  \\
		2024 May 01 & 60431.21 & JS & REOSC & 20.4  \\
            2024 Jul 03 & 60494.90 & Clay & LDSS-3 & 83.9  \\
    	\hline
	\end{tabular}
\end{table}

\section{Observational Properties}
\label{sec:prop}

We take the explosion epoch as the midpoint between the last non-detection and the first detection with an uncertainty equal to half the interval between those epochs, at MJD = 60410.795 ± 0.345. We adopt the redshift given by \citet{koribalski04} of 0.0024 to correct the spectra and the light curve phases. 

\subsection{Light curves and color evolution}

\begin{figure*}
    \centering
    \includegraphics[scale=1]{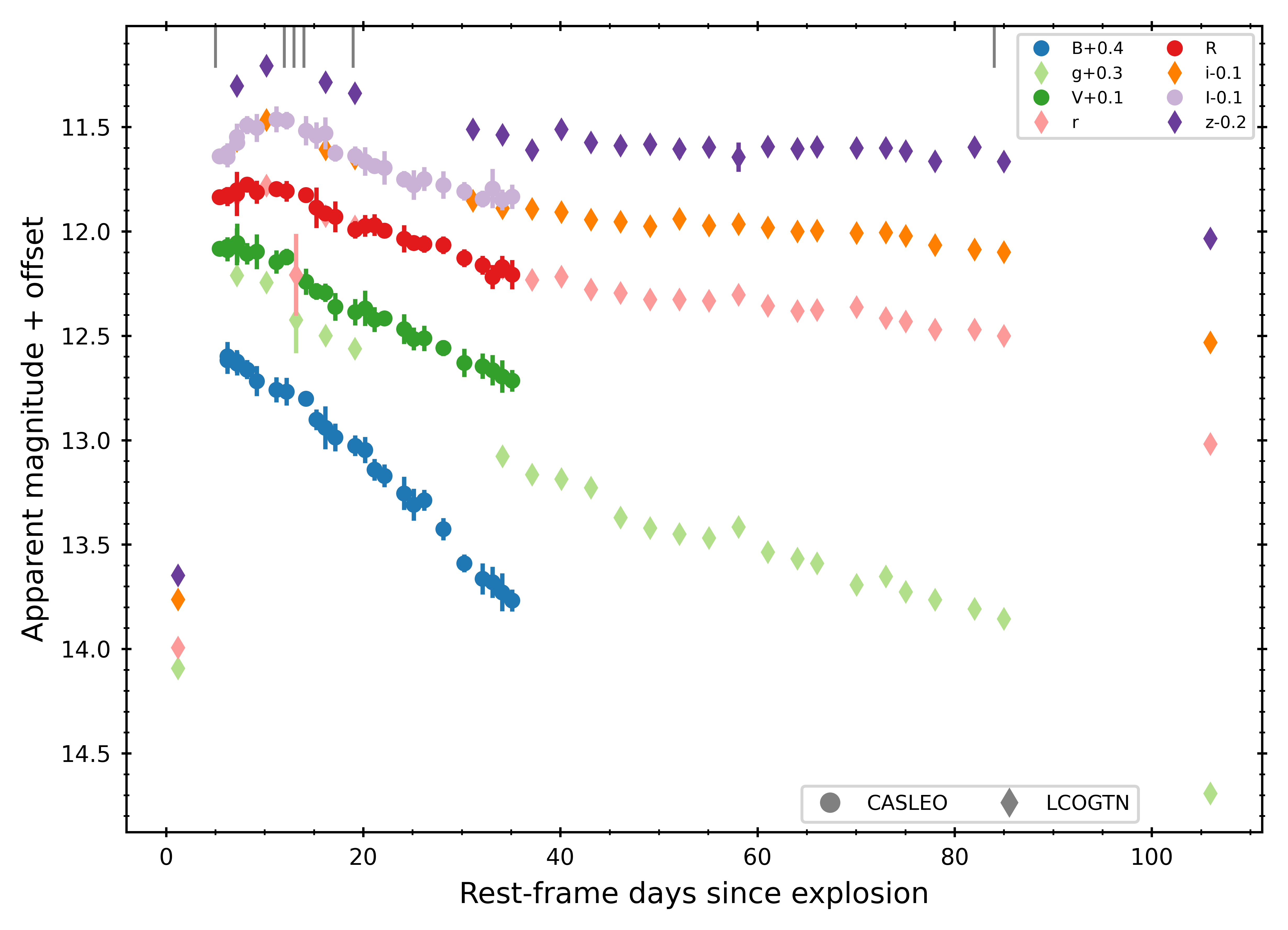}
    \caption{Observed light curves of SN~2024ggi. For clarity, the light curves are shifted by the offsets indicated in the upper legend. Different instruments are indicated as different markers. Rest-frame epochs of optical spectra are marked as gray lines along the top axis.
    }
    \label{fig:lcs}
\end{figure*}

The $BVRI$ and $griz$ light curves are presented in Figure \ref{fig:lcs}. The photometry from LCOGTN has been transformed from AB to Vega photometric system for a better display. Table \ref{tab:phot_griz} shows the original measurements in the AB system. We fitted the light curves (LC) corrected by extinction (see Section~\ref{sec:bol}) using a low-order polynomial in order to get peak magnitudes and rise times. This was done only on $griz$ since these bands are better sampled and have enough coverage during the rise. The results are listed in Table \ref{tab:phot_param}. We obtained an absolute peak magnitude in the $r$-band of $M_{r}$= $-17.73$ mag at MJD = 60419.63, giving a rise time of $t_{r}$ = 8.8 days. The uncertainty in the absolute peak magnitude is the result of adding in quadrature the fitting error, the uncertainty in the peak phase, and the uncertainty in the distance. The uncertainty in the rise time is taken as the uncertainty in the explosion epoch, since it is the main source of error. 

The rise times are longer the redder the band is, indicating that the peak is due to a decrease in temperature, corresponding to the well-known cooling phase prior to the recombination of SNeII. The plateau lasts around 90 days in $r$, $i$ and $z$ bands, while there is a minor decline in the $g$-band. 

    \begin{table}
      \caption{Light curve properties of SN~2024ggi}
         \label{tab:phot_param}
          \centering
         \begin{threeparttable} 
         \begin{tabular}{lp{0.1\textwidth}cp{0.1\textwidth}cp{0.1\textwidth}cp{0.1\textwidth}}
            \hline
            Filter & $\mathrm{MJD_{max}}$ & $t_{rise}$\tnote{a} [days] & $M_{max}$ [mag] \\
            \hline
             g & 60418.00 & 7.2 & -$17.87 \pm 0.02$   \\
             r & 60419.63 & 8.8 & -$17.73 \pm 0.01$  \\
             i & 60422.93 & 12.14 & -$17.84 \pm 0.01$  \\
             z & 60422.70 & 11.9 & -$17.40 \pm 0.01$ \\   
            \hline
        \end{tabular}
        \begin{tablenotes}
    \item[a] The uncertainty in the rise times is set to 0.3 days, corresponding to the uncertainty in the explosion epoch, which is the dominant source of error.
    \end{tablenotes}
    	\end{threeparttable}
    \end{table}

\subsection{Spectral properties} \label{sec:spec}

\begin{figure*}
    \centering
    \includegraphics[width=.8\linewidth]{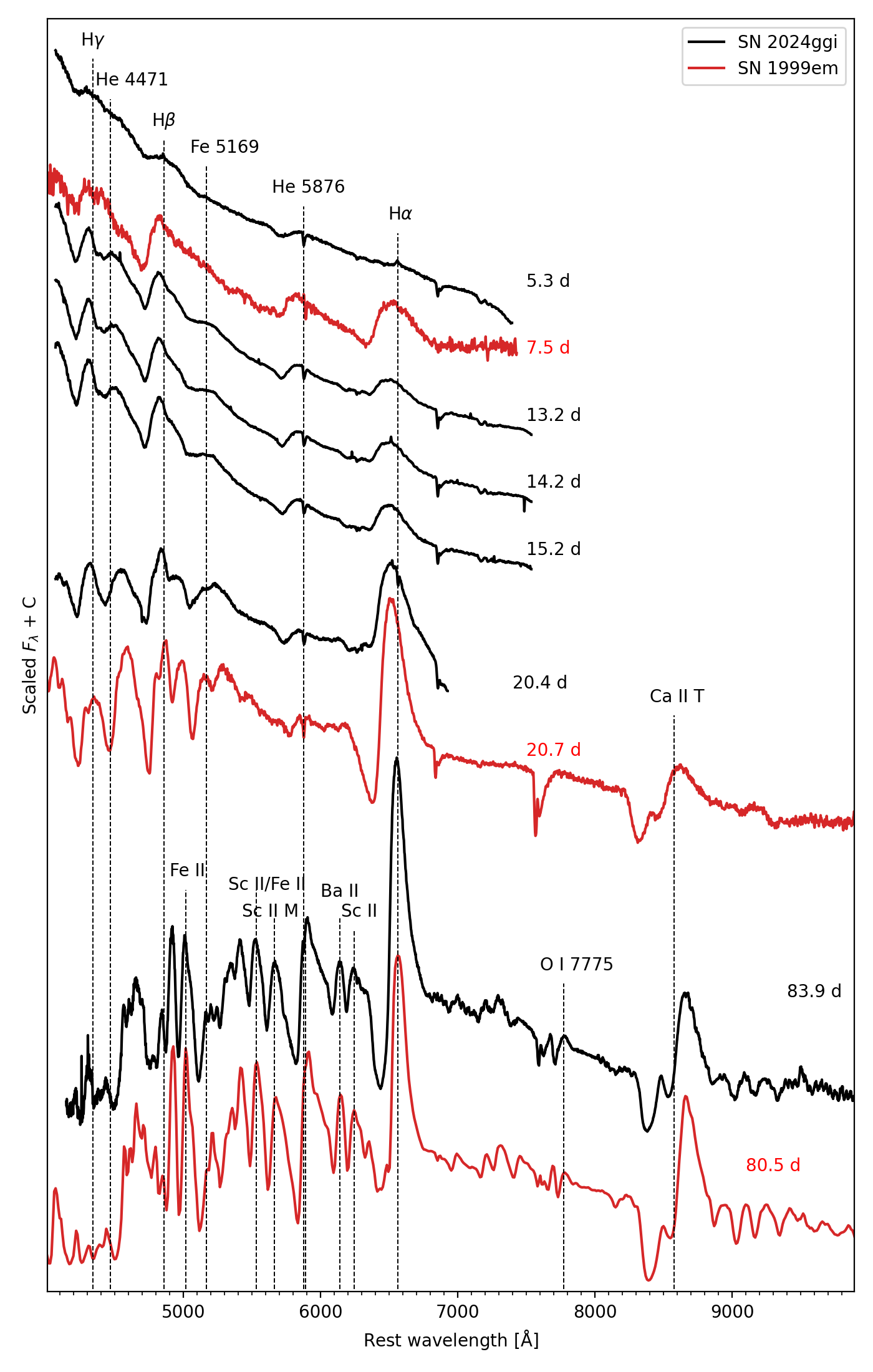}
    \caption{Spectral sequence of SN~2024ggi taken with the JS and Magellan Clay Telescopes, marked in black. Main absorption lines are marked with dashed lines at the rest wavelength. Spectra on SN 1999em is shown for comparison in red; epochs are referred to the explosion date derived by \cite{elmhamdi03}}
    \label{fig:spec}
\end{figure*}

The spectroscopic data of SN 2024ggi covering the phases from 5 to 84 days after the explosion is presented in Figure \ref{fig:spec}, compared with the standard Type II plateau SN~1999em. The rest-wavelength positions of the main features are marked with dashed lines. Our spectra cover well the interval between 5 and 19 days, and then there is a gap until 84 days. During the evolution, the continuum becomes less blue with time, consistent with the color evolution derived from the light curves. As noted by \citet{jacobsongalan24} and \citet{shrestha24}, in the spectrum at five days after explosion, which is our first spectrum, there is no evidence of high ionization lines. In the first spectrum, weak, broad absorption features of H$\beta$, He I $\lambda$ 4471, and He I $\lambda$ 5876 are present. The SN evolves slowly during the first 20 days, developing typical P-Cygni profiles. At 12 days, the H$\alpha$ and Fe I $\lambda$ 5169 profiles are clearly visible. The velocity of the H$\alpha$ line, measured from the absorption minimum, goes from $\approx-9500$ to $\approx-9000$ \kms~between 5 and 19 days, while Fe I $\lambda$ 5169 velocity varies from $\approx-7000$ to $\approx-6500$ \kms. 

The spectra of SN~2024ggi matches those of typical SNeII, as noticed by the comparison with SN~1999em. However, during the first 20 days of its evolution, the H$\alpha$ absorption is weaker and less pronounced than that observed in SN~1999em. This weak H$\alpha$ profile may be associated with the presence of ejecta-CSM interaction \citep{hillier19} and it is also linked to brighter and more rapidly declining SNeII \citep{gutierrez14}. Between 12 and 19 days, the spectra of SN~2024ggi exhibit the "Cachito" feature, an absorption component on the blue side of H$\alpha$ \citep{gutierrez17b}, which is also seen in the 21-day spectrum of SN~1999em. For SN~1999em, this feature had been proposed to be due to high velocity structures in the expanding ejecta of the SN \citep{baron00,leonard02}. Since then, it is been identified in different SNe as either high velocity H$\alpha$ or Si II~6355 \AA~\citep{pastorello06}. When interpreted as high velocity H$\alpha$, this feature may be associated with interaction between the ejecta and CSM \citep{chugai07}. \citet{gutierrez14} proposed that when "Cachito" is detected within 30 days post-explosion, it is more likely to be associated with SiII, whereas detections at later epochs are linked to high-velocity H$\alpha$. Since the "Cachito" feature in SN2024ggi is observed within this early time frame, it is likely associated with Si~II.

At 84 days, the SN has evolved considerably, decreasing its blue continuum and developing features corresponding to Fe II, Sc II, H and O I. The H$\alpha$ profile becomes strong, followed by the calcium NIR triplet feature. This last spectrum covers the wavelength range beyond 7200 \AA~and until 9700 \AA.~
At this phase, the H$\alpha$ absorption has shifted to $\approx-5900$ \kms, and the Fe II $\lambda$ 5169 to $\approx-3200$\kms. All the features described here are marked in Figure \ref{fig:spec}. Note that at 84 days, the "Cachito" feature had disappeared.

\subsection{Bolometric evolution}
\label{sec:bol}
We first calculate the bolometric luminosity of SN~2024ggi. To have accurate constraints on the CSM properties, early observations are needed. Our data set spans from 5 days to 35 days from explosion in $BVRI$ and from 2 to 106 in $griz$. Since we only made one observation in the first 5 days after explosion, which are crucial to determine the CSM properties, we add high cadence early photometry of SN~2024ggi published by \citet{shrestha24} to our observations. This data set is composed by ultraviolet (UV) and optical observations in the filters $UVW2$, $UVM2$, $UVW1$, $U$, $B$, $g$, $V$, $r$, and $i$, from the first day until 21 days after explosion.  

Although the observations are heavily sampled, magnitude values are sometimes missing for certain filters at a given epoch. To have the light curves with the same cadence across all filters, we linearly interpolate them. The gap between consecutive epochs to interpolate was always less than 2.5 days.

The next step is to correct the observed magnitudes by extinction. Regarding the Milky-Way (MW) extinction, the recalibrated dust maps of \citet{schlegel98} yield a value of $E(B-V)_{MW}=0.07$ mag \citep{schlafly11} from the NASA Extragalactic Database (NED\footnote{\url{https://ned.ipac.caltech.edu/}}), considering an extinction law from \citet{cardelli89} with $R_{V}=3.1$. \citet{thallis24} find three intervening galactic clouds in the line of sight to the SN using high-resolution spectroscopy, inferring a total extinction of $E(B-V)_{MW}=0.12 \pm 0.02$ mag. This value does not compare well with the extinction from the recalibrated maps of \citet{schlafly11}. As noted by \citet{thallis24}, the dust extinction map of \citet{schlegel98} is less accurate when multiple dust clouds with different temperatures are encountered, which may explain the discrepancy. Additionally, several estimates have been made for the host galaxy component of the extinction for this SN. \citet{jacobsongalan24} inferred an $E(B-V)_{host}= 0.084 \pm 0.018$ mag by calculating the Na I D1 and D2 equivalent widths (EW) from high resolution spectra, and using the calibrations from \citet{stritzinger18}. Similarly, \citet{thallis24} calculated the host extinction to be $E(B-V)_{host}= 0.036 \pm 0.007$ mag, and \citet{shrestha24} measured $E(B-V)_{host}= 0.034 \pm 0.020$ mag, both using the calibrations from \citet{poznanski12}. 
In summary, \citet{jacobsongalan24}, \citet{thallis24}, and \citet{shrestha24} calculate a total $E(B-V)_{tot}$ of $0.154$, $0.16$, and $0.154$ mag respectively. We assume $E(B-V)_{tot} = 0.16$ mag.

After correcting the interpolated light curves by extinction, we converted them to monochromatic fluxes at the effective wavelength of each band. Then, we integrated the monochromatic fluxes along wavelength for each epoch, obtaining the quasi-bolometric flux ($F_{\mathrm{qbol}}$). To account for the flux outside the observed wavelength range, we assumed that at early epochs, the SN emission is well represented by a black body (BB) distribution. We then extrapolated the UV and infrared (IR) to obtain the unobserved UV and IR flux ($F_{\mathrm{UV}}$ and $F_{\mathrm{IR}}$, respectively), by fitting a BB to the spectral energy distributions at each epoch. Blackbody fits were restricted to observational epochs with at least four bands, whether observed or interpolated. Then, the total bolometric flux was calculated as $F_{\mathrm{bol}}$ = $F_{\mathrm{UV}}$ + $F_{\mathrm{qbol}}$ + $F_{\mathrm{IR}}$, and converted to luminosity assuming the distance stated in Section \ref{sec:intro}. The uncertainty in the luminosity was estimated by considering uncertainties in the photometry, distance, and the estimated errors of the extrapolated fluxes.

Our high-cadence data extend up to $\sim$85 days post-explosion, then there is a 12 day gap before our last observation at 106 days post-explosion. Since this gap is too long to extrapolate,
to calculate the bolometric light curve beyond 85 days we used public photometry available in the B and V bands from the American Association of Variable Star Observers (AAVSO) Web page\footnote{\url{https://www.aavso.org/}} \citep{datos_aavso}. Around $\sim$250 photometric measurements in the $B$-band and $\sim$400 in the $V$-band are available contributed by different observers globally for SN~2024ggi, covering 125 days of the SN evolution. We adopted the mean magnitudes in daily bins after rejecting discrepant observations, beginning from 85 days post-explosion where our original data concluded. We corrected the magnitudes by extinction, and then we used the (B-V) color-based bolometric corrections from \citet{martinez22} to derive the bolometric magnitudes. The bolometric luminosities were then calculated using the distance adopted in Section \ref{sec:intro}. The uncertainties were calculated considering uncertainties in the photometry and in the bolometric corrections. 

The complete bolometric LC is shown in Figure \ref{fig:models}. We determined the bolometric magnitude at maximum to be $M_{max}=-18.920 \pm 0.001$ mag, and calculated the morphological bolometric LC parameters as defined by \citet{martinez22} (see their Figure 8 and \citealt{anderson14}). These parameters essentially characterize the bolometric magnitudes in different parts of the light curve, decline rates, and duration of the different phases.  In summary each parameter is defined as follows: $s1$, $s2$, and $s3$ are the decline rates in magnitudes per 100 days during the cooling phase, the plateau phase, and the radioactive tail phase, respectively. The parameter $t_{trans}$ corresponds to the epoch of transition between the cooling decline and the plateau decline. $optd$, $p_{d}$, and ${C_{d}}$ correspond to the duration of the optically-thick, plateau, and cooling phases, respectively. Finally, $M_{bol,end}$ and $M_{bol,tail}$ are the bolometric magnitudes measured 30 days before and after $t_{PT}$, respectively, where $t_{PT}$ is equivalent to $optd$. Due to the lack of data during the radioactive tail phase at the time of these calculations, we do not include values for $M_{tail}$ or the $s_{3}$ decline rate. The results are listed on Table \ref{tab:bol_param}, together with the comparison of the same parameters for SN~2023ixf from \citet{bersten24}. Additionally, we included in Table \ref{tab:bol_param} the parameters calculated by \citet{martinez22} using a large sample of SNeII from the Carnegie Supernova Project-I \citep[CSP-I,][]{hamuy06}.
 
We find that most parameters of SN~2024ggi, similar to SN~2023ixf, fall within 1$\sigma$ of the comparison distributions, suggesting it is a typical Type II supernova. However, we note some minor deviations: SN 2024ggi exhibits a longer plateau duration and declines faster in the cooling phase than average, contrary to which was obtained for SN 2023ixf, which exhibited a shorter plateau duration, compared to the distribution of SNeII. A longer plateau duration suggests a more massive progenitor than the bulk of SNeII \cite{martinez22b} (see Section \ref{sec:hydro}).

    \begin{table}
      \caption{Bolometric light curve parameters of SN~2024ggi}
         \label{tab:bol_param}
          \centering
        \begin{tabular}{lccc}
            \hline
            Parameter & SN~2024ggi & SN~2023ixf & CSP-I \\
            \hline
             $M_{bol,end}$~[mag] & $-16.68(0.001)$ & -$17.18(0.06)$  & -$16.2(0.6)$  \\
             $s_{1}$~[mag/100d] & $7.12(0.10)$ &  $5.53(0.91)$ & $4.59(2.84)$  \\
             $s_{2}$~[mag/100d] & $0.55(0.02)$ &  $1.84(0.56)$ & $0.81(0.91)$   \\   
             $C_{d}$~[d] & $26.21(1.75)$ &  $29.66(5.31)$ & $26.9(4.3)$  \\
             $p_{d}$~[d] & $93.23(1.41)$ &  $53.42(5.23)$ & $75.0(26.2)$   \\
             $optd$~[d] & $119.44(0.34)$ & $83.08(0.08)$ & $ 104.3(19.3)$   \\  
            \hline
        \end{tabular}
    \end{table}

\section{Hydrodynamical modeling}
\label{sec:hydro}

We aimed at inferring the progenitor and CSM properties by comparing the bolometric LC and velocity evolution to models computed using the one-dimensional (1D) Lagrangian local thermodynamic equilibrium (LTE) radiation hydrodynamics code from \citet{bersten11}. Given that moderate CSM structures do not significantly influence bolometric LCs of SNe II at times $\gtrsim 30$~d \citep{morozova18,martinez22b}, it is practical to divide the SN 2024ggi modeling into two steps. 

First, in Section \ref{sec:global} we focused on inferring the global parameters such as the explosion energy ($E_\mathrm{{exp}}$), the progenitor mass and radius, and the nickel mass ($\mathrm{M_{^{56}Ni}}$) and its mixing ($\mathrm{mix(^{56}Ni)}$\footnote{The $\mathrm{^{56}Ni}$ mixing is measured as a percentage of the total pre-SN mass.}) by matching the plateau luminosity and duration, as well as the Fe II line velocities with our models. 

Then, in Section \ref{sec:CSM} we focused on deriving the CSM parameters such as the CSM extension ($R_\mathrm{{CSM}}$) and mass loss rate ($\dot M$) for different wind prescriptions (steady and accelerated). The CSM was artificially included in the outermost regions of our progenitor models. We then matched the cooling phase luminosity, duration, and steepness, as well as the line velocities, with our models. In this work, we adopted a cooling phase duration of $Cd=26$~d, derived from the bolometric LC (see Section \ref{sec:bol}), to distinguish the CSM interaction-dominated phase from the rest of the evolution. Lastly, a complete model was calculated using the parameter set derived from the two-step modeling.

\subsection{Global parameter modeling}
\label{sec:global}

Our hydrodynamical code requires progenitor models at the time of core collapse in order to initialize the explosion. For this work we utilized a pre-SN model grid calculated by \citet{nomoto88}, which comprises an RSG set with zero-age main sequence (ZAMS) masses of 13, 15, 18, 20 and 25 $M_{\odot}$. For simplicity, we will refer to these progenitor models by their ZAMS masses prefixed with the letter M (e.g., M13). The main properties of our pre-SN models are listed in Table \ref{tab:presn}. The explosion is then simulated by depositing some energy in the form of a thermal bomb at a mass coordinate where the pre-SN structure is assumed to collapse into a compact remnant ($M_\mathrm{core}$), and is thus removed from our calculations.

\begin{table}
    \caption{Properties of the pre-SN model grid calculated by \citet{nomoto88} used in this work. From left to right: model name, ZAMS mass ($M_\mathrm{ZAMS}$), pre-SN mass ($M_\mathrm{pre-SN}$), pre-SN radius ($R_\mathrm{pre-SN}$), compact core remnant mass ($M_\mathrm{core}$), H mass ($M_\mathrm{H}$) and ejecta mass ($M_\mathrm{ej}$).}
    \label{tab:presn}
    \centering
    \begin{tabular}{lcccccc}
        \hline
        Model & $M_\mathrm{ZAMS}$ & $M_\mathrm{pre-SN}$ & $R_\mathrm{pre-SN}$ & $M_\mathrm{core}$ & $M_\mathrm{H}$ & $M_\mathrm{ej}$ \\
        \hline
        M13 & 13 & 12.73 & 576  & 1.6 & 6.17 & 11.13 \\
        M15 & 15 & 14.11 & 517  & 1.7 & 6.77 & 12.41 \\
        M18 & 18 & 16.74 & 729  & 1.7 & 7.51 & 15.04 \\
        M20 & 20 & 18.35 & 812  & 1.8 & 7.93 & 16.55 \\
        M25 & 25 & 21.69 & 1234 & 2.0 & 8.44 & 19.69 \\
        \hline
    \end{tabular}
\end{table}

A set of hydrodynamical models was generated by varying several physical parameters for each of our pre-SN models. We explored different values of $E_\mathrm{exp}$ between $0.5-1.7~\mathrm{foe}$ ($1~\mathrm{foe} = 10^{51}~\mathrm{erg}$), $\mathrm{M_{^{56}Ni}}$ between $0.01-0.06~M_\mathrm{\odot}$ and $\mathrm{mix(^{56}Ni)}$ between $10-80~\%$. Then, we visually compared the model set with the bolometric LC and Fe II line velocities of SN 2024ggi derived in Sections \ref{sec:spec} and \ref{sec:bol} to select our preferred model.

\begin{figure*}
    \centering
    \includegraphics[page=1,width=0.49\textwidth]{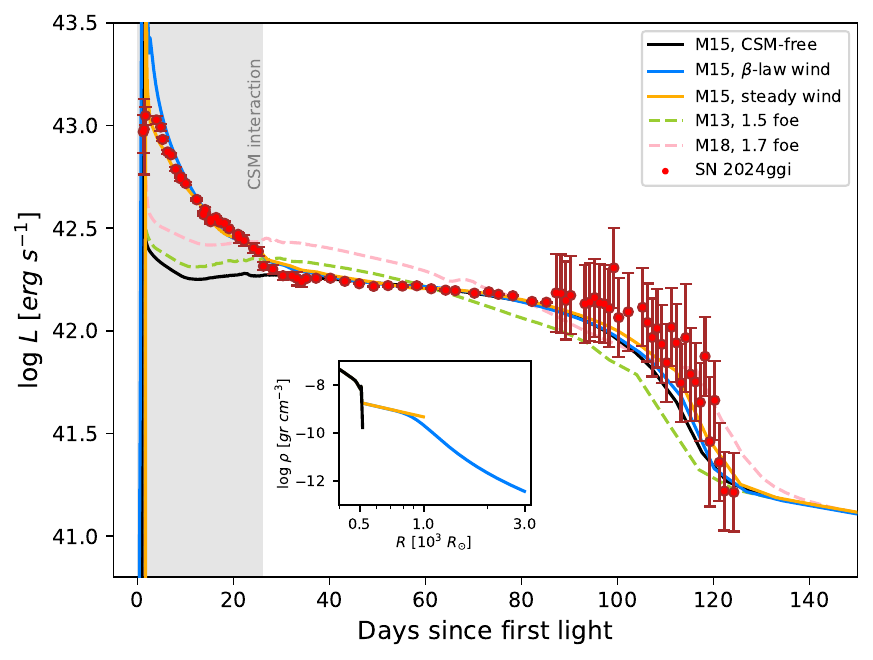}
    \includegraphics[page=2,width=0.48\textwidth]{figures/SN2024ggi_models_v5.pdf}
    \caption{Observations of SN~2024ggi (points) compared with hydrodynamical models (lines). Left panel: Bolometric light curve. The inset panel shows density profiles of the models with no CSM (black), steady wind (yellow), and accelerated wind (blue). Right panel: Photospheric velocity evolution. The shaded gray area marks the approximate time frame where the emission is dominated mainly by CSM interaction, for times $t \lesssim Cd = 26~d$. The uncertainty in the velocities is taken as $\approx-500$ \kms and it takes into account the resolution of the spectra and the error in the measurements.}
    \label{fig:models}
\end{figure*}

To choose an adequate progenitor model, we use the well-known fact that for a given pre-SN model (with a fixed mass and radius), $E_\mathrm{exp}$ is the only parameter that can modify the ejecta expansion velocities \citep{Kasen2009,Dessart2013,Bersten2013}. Thus, our initial exploration focused on finding appropriate $E_\mathrm{exp}$ values to reproduce the Fe II velocities of SN 2024ggi across our pre-SN model grid. From this analysis, we found that only the M13, M15 and M18 models could reproduce the Fe II line velocities using $E_\mathrm{exp}$ values of $1.5$, $1.3$ and $1.7~\mathrm{foe}$, respectively, and also provide a reasonable LC of SN 2024ggi. For more massive models, no solution was found.

Subsequently, we focused on refining the bolometric LC match through an exploration of the nickel mass and distribution ($\mathrm{M_{^{56}Ni}}$ and $\mathrm{mix(^{56}Ni)}$), the remaining free parameters available. From this exploration we found that M15 models provide an overall better agreement with the observations of SN 2024ggi.

In Figure \ref{fig:models} we present our preferred solution in solid black line, which corresponds to the M15 model with an explosion energy of $E_\mathrm{exp}=1.3~\mathrm{foe}$, a $^{56}$Ni mass of $\mathrm{M_{^{56}Ni}} = 0.035~M_{\odot}$ with $\mathrm{mix(^{56}Ni)} = 30~\%$. We also present two additional models, M13 (dashed green line) and M18 (dashed pink line), as a comparison. Although these models provide a comparable match to the Fe II velocities as our preferred M15 model, the same is not true for the LC of SN 2024ggi. As can be seen in Figure \ref{fig:models}, both models have a steeper plateau decline rate than observed, overestimate the plateau luminosity early on, and produce a shorter (M13) or longer (M18) plateau duration depending on the model. Therefore, these models produce a poorer match to the data than our preferred M15 model. However, we cannot rule out some intermediate models from our analysis.

The M15 model provides a good representation of the plateau, the transition and the onset of the radioactive tail. However, we posit that the derived nickel properties should be taken with caution due to the relatively large uncertainties starting at $t \sim 87$~d, combined with the lack of observations beyond $t \sim 124$~d after the explosion. Nevertheless, given the strong correlation between radioactive tail luminosity and $\mathrm{^{56}Ni}$ mass \citep{Martinez2022c}, it is possible to rule out values of $M_\mathrm{{^{56}Ni}} \gtrsim 0.035~M_{\odot}$, as they produce radioactive tail luminosities exceeding the faintest observed transition luminosity. We also note that our choice of progenitor and $E_\mathrm{exp}$ parameters is not altered by the exploration of nickel parameters, since the latter do not affect the expansion velocities and have a comparatively small influence on the plateau characteristics.

From our exploration, we conclude that the M15 model with an explosion energy of $E_\mathrm{exp}=1.3~\mathrm{foe}$, a $^{56}$Ni mass of $\mathrm{M_{^{56}Ni}} = 0.035~M_{\odot}$ with $\mathrm{mix(^{56}Ni)} = 30~\%$ is a model that represents well the Fe II velocities and the bolometric LC of SN 2024ggi at times $t \gtrsim 26~\mathrm{d}$. This model is presented in solid black line in Figure \ref{fig:models}. Although our analysis is based on visual comparisons, we deem our choice of the optimal physical parameters to be well justified within the assumptions of our modeling. A more refined statistical analysis is beyond the scope of this study.

\subsection{CSM parameters modeling} 
\label{sec:CSM}

The models presented in Section \ref{sec:global} fail to reproduce the early observations since they underestimate the bolometric luminosity up to $t\sim26~\mathrm{d}$. This discrepancy has been attributed to the effect of the interaction between the ejecta and an existing CSM. It has been established that the incorporation of a CSM distribution at the outermost layers of the pre-SN structure increases the luminosity of the resulting model during the cooling phase, thus improving the early-time modeling \citep{Moriya2011,morozova18,Englert2020}. The presence of CSM also lowers the maximum photospheric velocity and halts the velocity decline during the cooling phase, which can help constrain the plausible CSM configurations. The existence of a CSM structure in SN 2024ggi is further supported by the presence of flash features in the early-time spectra \citep{hoogendam24,jacobsongalan24,thallis24,chen24,shrestha24}.

On that basis, we modified the density profile of the M15 progenitor model by attaching a CSM distribution before simulating the explosion. We use the same explosion parameters as those of our preferred model presented in Section \ref{sec:global}. Only the CSM properties are explored, which mainly affect the LC and expansion velocities for $t \lesssim 26~\mathrm{d}$. We note that different CSM configurations introduce slight variations during the transition to the radioactive tail. However, these differences are too small to warrant a re-evaluation of the model parameters derived in Section \ref{sec:global}.

In this section we present two different scenarios: a steady wind distribution ($\rho \propto r^{-2}$) and an accelerated wind distribution. In both cases, a set of models was explored and visually compared with the SN 2024ggi data. In the following, we present and discuss the best models found within our exploration. However, it must be noted that we cannot rule out other possible solutions given the qualitative nature of our analysis and the well-known degeneracies between CSM parameters \citep{Dessart23, Khatami24}. To refine the parameter exploration, a statistical study with a broader parameter grid needs to be performed, which is left for future work.
  
For the steady wind scenario, the wind velocity was fixed at $v_w$ = 77~km\,s$^{-1}$, as measured by \cite{thallis24}, and different CSM extensions and mass loss rates were explored. The preferred steady-wind model is shown in solid orange line in Figure \ref{fig:models}, and it greatly improves the early LC and expansion velocities modeling as a result of the inclusion of this CSM. Said model has an extension of $R_\mathrm{{CSM}} = 1000~R_{\odot}$ and a mass loss rate of $\dot M = 3.6~M_{\odot}$\,yr$^{-1}$, corresponding to a CSM mass of $M_\mathrm{{CSM}} = 0.5~M_{\odot}$. The inferred mass loss rate is considerably higher than the typical range for SNe II-P, suggesting an enhanced mass loss event during the last $\sim 50$~d before the explosion \citep{Morozova2017}. 

We also examined whether this model was able to reproduce the duration of the flash features in the observed spectra. Following \citet{dessart17}, the narrow lines last as long as the shock is placed within a slow-moving optically thick material (i.e. until the shock goes through the SN photosphere). In our model we found that the flash features should disappear $\sim0.2~\mathrm{d}$ after shock breakout. This duration is an order of magnitude lower than the estimated value of $3.8 \pm 1.6~\mathrm{d}$ for SN 2024ggi \citealt{jacobsongalan24} and could be a plausible reason to consider the steady wind model less favorably.

For the accelerated wind scenario, we followed the wind velocity prescription given by \citet{moriya18} which takes the form of the $\beta$ velocity law given below:

\begin{equation}
    v_{w}(r) = v_0 + (v_\infty - v_0)(1-R_0/r)^\beta,
\end{equation}

\noindent Where $v_0$ is the initial wind velocity (0.1~km\,s$^{-1}$), $v_\infty$ is the terminal wind velocity (77~km\,s$^{-1}$, \citealt{thallis24}), $R_0$ is the radial coordinate where the CSM is attached to the progenitor model, and $\beta$ is the wind acceleration parameter \citep{lamers99}.

We then explored different CSM extensions, mass loss rates and wind acceleration parameters, and compared the resulting model grid with the early-time bolometric LC and line velocities. The preferred accelerated wind model has an extension of $R_\mathrm{{CSM}} = 3000~R_{\odot}$, a mass loss rate of $\dot M = 4.6 \times 10^{-3}~M_{\odot}$\,yr$^{-1}$ and a wind acceleration parameter of $\beta = 9$, corresponding to a CSM mass of $M_\mathrm{{CSM}} = 0.55~M_{\odot}$. This model, shown in solid blue line in Figure \ref{fig:models}, greatly improves the modeling of early-time observations compared to the CSM-free model. It produces a bolometric LC similar to the steady wind model, albeit more luminous and with a steeper decline rate during the first $\sim10~\mathrm{d}$ of evolution. Likewise, the photospheric velocity evolution of both CSM models is comparable, although the accelerated wind model yields slightly lower velocities during the first $\sim15~\mathrm{d}$. Since Fe II velocity measurements before $\mathrm{t} \lesssim 15~\mathrm{d}$ are lacking, we cannot further constrain the CSM properties of SN 2024ggi. 

We note that the optimal wind acceleration parameter found in our exploration is higher than the typical range for normal RSGs ($1-5$, \citealt{moriya18}). This would be consistent with an enhanced mass-loss event scenario prior to the explosion. We also examined the duration of the flash features, and found that they should last for $\sim1.2~\mathrm{d}$ after shock breakout. This is an improvement over the relatively short-lived prediction in our preferred steady wind model, and closer to, though still shorter than the observed duration in SN 2024ggi ($3.8 \pm 1.6~\mathrm{d}$, \citealt{jacobsongalan24}).

Despite the accelerated wind scenario producing a $ \dot M$ value three orders of magnitude lower than the steady wind model, we find that the total $M_\mathrm{CSM}$ remains roughly similar between the two cases. This consistency in the inferred $M_\mathrm{CSM}$ is noteworthy, as it suggests that a similar amount of material was needed in both scenarios to decelerate the shock wave and thereby produce lower expansion velocities. On the other hand, the mass loss rate is associated with the late evolutionary history of the progenitor star, and thus remains largely unconstrained despite recent efforts \citep{Quataert2012,Woosley2015,Fuller2024}. 

The density profiles of the steady and accelerated wind models are shown in the inset of Figure \ref{fig:models}. Both CSM profiles exhibit similar density and steepness up to a radius of $R \simeq 1000~R_{\odot}$, indicating the presence of a dense and compact CSM core. Beyond this extension, in the range of $R \simeq 1000 - 3000~R_{\odot}$, the accelerated wind model shows a sharp drop in density forming a low-density tail. This configuration ---an inner dense and compact core coupled with an outer light and extended tail--- resembles the two-component CSM distribution proposed in recent studies \citep{Chugai2022,jacobsongalan23,Zimmerman2024}. The two-component CSM models offer a promising pathway to explain the CSM mass required to reproduce the early-time bolometric light curve while providing more realistic mass-loss scenarios. Therefore, we consider the accelerated wind model to be the more reasonable prescription for the CSM structure of SN 2024ggi, which in turn provides a more credible timescale for the duration of the flash features as discussed above.

\section{Conclusions}
\label{sec:conclusions}

In this work we present optical photometric and spectroscopic observations of the Type II SN~2024ggi, spanning from 2 to 106 days after explosion. Similar to SN~2023ixf, SN~2024ggi is among the closest supernovae of the decade, providing a unique opportunity to constrain the progenitor properties of Type II supernovae. The analysis of the bolometric LC suggests that SN~2024ggi is a typical Type II SN. Nevertheless, it shows a longer plateau duration and a faster decline in the cooling phase, compared to a distribution of SNeII. 
We presented the first hydrodynamical modeling of the bolometric LC and photospheric velocity evolution of SN~2024ggi, using the full extent of the plateau phase. Our results suggest that SN~2024ggi originated from the explosion of a star with a ZAMS mass of $15~M_{\odot}$, an explosion energy of $1.3 \times 10^{51}$\,erg, a $\mathrm{^{56}Ni}$ production $\lesssim$ 0.035 $M_{\odot}$ and a relatively moderate $\mathrm{^{56}Ni}$ mixing of $30\%$. The exploded RSG star at the final stage of its evolution had a mass of 14 $M_{\odot}$, and radius of 516 $R_{\odot}$. 

To characterize the CSM around the progenitor star, we modeled the early phases of the explosion by modifying the outermost density profile considering two different scenarios: steady winds and accelerated winds. In the steady wind case, the preferred model suggests a CSM extension of $1000~\mathrm{R_{\odot}}$ ($7 \times 10^{13}$cm) with a mass-loss rate of $3.6~\mathrm{M_{\odot}}$ yr$^{-1}$, corresponding to a total CSM mass of $0.5~\mathrm{M_{\odot}}$. This model predicts a maximum flash features duration of $0.2~\mathrm{d}$. For the accelerated wind case, the preferred model points to a CSM extension of 3000~$\mathrm{R_{\odot}}$ ($2.1 \times 10^{14}$cm) with a mass-loss rate of $4 \times 10^{-3}\mathrm{M_{\odot}}$ yr$^{-1}$, and an acceleration parameter of $\beta=9$, resulting in a similar CSM mass of $0.55~\mathrm{M_{\odot}}$. In this case, the duration of the flash features is extended until $1.2~\mathrm{d}$ which is closer in duration to the observations. While both models reproduce the bolometric LC and expansion velocity evolution reasonably well, we consider the accelerated wind scenario to be more reasonable as it provides a lower mass-loss rate and a slightly better agreement with the duration of the flash features. 

There are several works in the literature analyzing early properties of SN~2024ggi, which allow us to compare our inferred parameters. Studies modeling early spectra of SN~2024ggi found a CSM confined to a range of $R_\mathrm{{CSM}}=2.7-5 \times 10^{14}$~cm (3900 - 7200~$\mathrm{R_{\odot}}$), formed from a progenitor with mass-loss rate in the range of ~$10^{-3}$-$10^{-2}$~$\mathrm{M_{\odot}}$~$\mathrm{yr^{-1}}$ \citep{jacobsongalan24,shrestha24,zhang24}. Additionally, \citet{chenTW24} presented a hydrodynamical model of the first $\sim$15 days of only the light curve information and found that the data are well-matched by a model with an explosion energy of $2\times10^{51}$ erg, a mass-loss rate of ~$10^{-3}$ $\mathrm{M_{\odot}}$~$\mathrm{yr^{-1}}$ (assuming an accelerated wind with $\beta$ = 4.0 and a terminal wind velocity of 10 km~s$^{-1}$), and a confined CSM with a radius of $R_\mathrm{{CSM}}=6\times10^{14}$~cm ($\sim$$8600~\mathrm{R_{\odot}}$), and  a mass of 0.4 $\mathrm{M_{\odot}}$. Despite the difference in methodology, we found that our estimations of the CSM parameters, except for $R_\mathrm{{CSM}}$, are in agreement with those calculated in the literature.

Studies analyzing the pre-explosion data of the SN site identified a RSG star with an estimated mass ranging from 13 to 17 $\mathrm{M_{\odot}}$ as a progenitor candidate of SN 2024ggi \citep{xiang24,chenTW24}. Furthermore, environmental studies of the SN site suggest a lower-mass progenitor, compatible with 10 $\mathrm{M_{\odot}}$ \citep{hong24}. Our findings align with the estimates derived from direct detections. Moreover, our analysis of the morphological parameters of the bolometric LC of SN~2024ggi reveals a longer plateau compared to SN~2023ixf and with a sample of SNeII from the CSP-I. This suggests that the progenitor of SN2024ggi was more massive than that of SN2023ixf and than the average progenitor mass in the CSP-I sample, in line with what we find in our hydrodynamic modeling.

The recent detection of two of the closest SNeII in the decade, SN~2023ixf and SN~2024ggi, highlights the importance of early, high-cadence observations in constraining the physics of both the explosion and the progenitor stars of SNeII. Continued monitoring of SN~2024ggi during the nebular phase and after its emission fades, to confirm the disappearance of the progenitor candidate, will provide critical insights into its nature.

\begin{acknowledgements}
This work is based on data acquired at Complejo Astronómico El Leoncito, operated under agreement between the Consejo Nacional de Investigaciones Científicas y Técnicas de la República Argentina and the National Universities of La Plata, Córdoba and San Juan. We thank the authorities of CASLEO, for the prompt response that made possible to obtain the data. This paper includes data gathered with the 6.5 meter Magellan Telescopes located at Las Campanas Observatory, Chile, and observations from the Las Cumbres Observatory global telescope network under the program allocated by the Chilean Telescope Allocation Committee (CNTAC), no: CN2024B-35. E.H. was supported from ANID, Beca Doctorado Folio \texttt{\#}21222163. R.C. acknowledges support from Gemini ANID ASTRO21-0036. M.O. acknowledges support from UNRN~PI2022~40B1039 grant. M.G. acknowledges support from ANID, Millennium Science Initiative, AIM23-0001.

\end{acknowledgements}

\bibliographystyle{aa} 
\bibliography{2024ggi}

\begin{appendix}
\appendix
\let\clearpage\relax
\section{Tables}
\label{ap:1}
    \begin{table*}
      \caption{Optical photometry of SN~2024ggi with HSH and JS Telescopes. Photometry is not corrected by Galactic nor host extinction.}
         \label{tab:phot}
          \centering
        \begin{tabular}{lcccc}
            \hline
            MJD & B & V & R & I \\
            \hline
            60416.23229 & $11.968 \pm	0.764$ & $11.982 \pm 0.021$ & $11.835 \pm 0.022$ & $11.740 \pm	0.033$ \\
            60417.02009 & $12.217 \pm  0.065$ & $11.973 \pm 0.044$ & $11.832 \pm 0.046$ & $11.744 \pm	0.049$  \\
            60417.04254 & $12.197 \pm  0.068$ & $11.989 \pm 0.053$ & $11.826 \pm 0.033$ & $11.723 \pm	0.045$  \\
            60418.00073 & $12.222 \pm  0.054$ & $11.973 \pm 0.088$ & $11.820 \pm 0.105$ & $11.647 \pm	0.064$  \\
            60418.02609 & $12.233 \pm  0.056$ & $11.955 \pm 0.092$ & $11.802 \pm 0.033$ & $11.675 \pm	0.038$  \\
            60419.01688 & $12.261 \pm  0.045$ & $12.006 \pm 0.051$ & $11.776 \pm 0.034$ & $11.591 \pm	0.043$  \\
            60420.01423 & $12.316 \pm  0.072$ & $11.997 \pm 0.083$ & $11.812 \pm 0.054$ & $11.604 \pm	0.066$  \\
            60421.99991 & $12.358 \pm  0.060$ & $12.046 \pm 0.056$ & $11.797 \pm 0.036$ & $11.563 \pm	0.062$  \\
            60423.05799 & $12.367 \pm  0.066$ & $12.022 \pm 0.039$ & $11.807 \pm 0.049$ & $11.569 \pm	0.042$  \\
            60424.01363 & $12.430 \pm  0.834$ & $12.114 \pm 0.464$ & $11.844 \pm 0.095$ & $11.598 \pm	0.082$  \\
            60425.02442 & $12.401 \pm  0.027$ & $12.141 \pm 0.062$ & $11.826 \pm 0.029$ & $11.617 \pm	0.070$  \\
            60426.06700 & $12.502 \pm  0.049$ & $12.186 \pm 0.040$ & $11.886 \pm 0.097$ & $11.641 \pm	0.063$  \\
            60426.99785 & $12.540 \pm  0.102$ & $12.193 \pm 0.044$ & $11.914 \pm 0.031$ & $11.630 \pm	0.076$  \\
            60428.00727 & $12.586 \pm  0.066$ & $12.261 \pm 0.066$ & $11.929 \pm 0.073$ & $11.724 \pm	0.040$  \\
            60430.02084 & $12.626 \pm  0.049$ & $12.286 \pm 0.062$ & $11.991 \pm 0.042$ & $11.738 \pm	0.045$  \\
            60431.02126 & $12.646 \pm  0.062$ & $12.268 \pm 0.084$ & $11.973 \pm 0.052$ & $11.765 \pm	0.067$  \\
            60431.99448 & $12.741 \pm  0.052$ & $12.322 \pm 0.059$ & $11.970 \pm 0.052$ & $11.787 \pm	0.033$  \\
            60433.00000 & $12.770 \pm  0.055$ & $12.316 \pm 0.022$ & $11.996 \pm 0.036$ & $11.795 \pm	0.080$  \\
            60434.99879 & $12.854 \pm  0.079$ & $12.367 \pm 0.071$ & $12.035 \pm 0.064$ & $11.850 \pm	0.039$  \\
            60435.98729 & $12.908 \pm  0.076$ & $12.415 \pm 0.054$ & $12.055 \pm 0.031$ & $11.878 \pm	0.070$  \\
            60437.04790 & $12.887 \pm  0.050$ & $12.412 \pm 0.061$ & $12.060 \pm 0.041$ & $11.849 \pm	0.056$  \\
            60438.99389 & $13.026 \pm  0.052$ & $12.458 \pm 0.038$ & $12.065 \pm 0.041$ & $11.877 \pm	0.066$  \\
            60441.12399 & $13.189 \pm  0.043$ & $12.529 \pm 0.067$ & $12.127 \pm 0.042$ & $11.908 \pm	0.044$  \\
            60442.98297 & $13.264 \pm  0.074$ & $12.545 \pm 0.060$ & $12.161 \pm 0.045$ & $11.944 \pm	0.039$  \\
            60443.98952 & $13.280 \pm  0.074$ & $12.564 \pm 0.072$ & $12.217 \pm 0.057$ & $11.894 \pm	0.094$  \\
            60444.98770 & $13.328 \pm  0.091$ & $12.594 \pm 0.077$ & $12.170 \pm 0.053$ & $11.949 \pm	0.049$  \\
            60445.97769 & $13.367 \pm  0.052$ & $12.615 \pm 0.051$ & $12.207 \pm 0.070$ & $11.934 \pm	0.058$  \\  
            \hline
        \end{tabular}
    \end{table*}

    \begin{table*}
      \caption{Optical photometry of SN~2024ggi with the LCOGTN-1m telescope. Photometry is not corrected by Galactic nor host extinction.}
         \label{tab:phot_griz}
          \centering
        \begin{tabular}{lcccc}
            \hline
            MJD & g & r & i & z \\
            \hline
            60412 & $13.713 \pm	0.01$   & $14.153 \pm 0.012$ & $14.213 \pm 0.012$ & $14.387 \pm	0.026$ \\
            60418 & $11.831 \pm 0.018 $ & $11.988 \pm 0.01$  & $12.016 \pm 0.012$ & $12.043 \pm	0.016$  \\
            60421 & $11.865 \pm 0.014 $ & $11.941 \pm 0.014$ & $11.916 \pm 0.012$ & $11.947 \pm	0.015$  \\
            60424 & $12.044 \pm 0.159 $ & $12.368 \pm 0.196$ & -  & -  \\
            60427 & $12.119 \pm 0.012 $ & $12.086 \pm 0.012$ & $12.057 \pm 0.011$ & $12.026 \pm	0.014$  \\
            60430 & $12.182 \pm 0.015 $ & $12.137 \pm 0.008$ & $12.101 \pm 0.010 $ & $12.079 \pm	0.013$  \\
            60442 & -    & - & $12.304 \pm 0.017$ & $12.252 \pm	0.018$  \\
            60445 & $12.697 \pm 0.011 $ & $12.336 \pm 0.006$ & $12.339 \pm 0.006$ & $12.278 \pm	0.011$  \\
            60448 & $12.785 \pm 0.021 $ & $12.392 \pm 0.027$ & $12.343 \pm 0.017$ & $12.35  \pm	0.036$  \\
            60451 & $12.806 \pm 0.019 $ & $12.377 \pm 0.015$ & $12.358 \pm 0.011$ & $12.251 \pm	0.015$  \\
            60454 & $12.848 \pm 0.03  $ & $12.439 \pm 0.009$ & $12.394 \pm 0.009$ & $12.314 \pm	0.017$  \\
            60457 & $12.99 \pm  0.006 $ & $12.455 \pm 0.005$ & $12.404 \pm 0.006$ & $12.329 \pm	0.011$  \\
            60460 & $13.041 \pm 0.007 $ & $12.486 \pm 0.005$ & $12.425 \pm 0.008$ & $12.323 \pm	0.01$  \\
            60463 & $13.069 \pm 0.009 $ & $12.486 \pm 0.009$ & $12.39  \pm 0.018$ & $12.346 \pm	0.043$  \\
            60466 & $13.088 \pm 0.008 $ & $12.493 \pm 0.006$ & $12.421 \pm 0.008$ & $12.337 \pm	0.014$  \\
            60469 & $13.036 \pm 0.032 $ & $12.464 \pm 0.012$ & $12.415 \pm 0.012$ & $12.384 \pm	0.07$  \\
            60472 & $13.156 \pm 0.007 $ & $12.516 \pm 0.005$ & $12.431 \pm 0.005$ & $12.334 \pm	0.008$  \\
            60475 & $13.187 \pm 0.01  $ & $12.541 \pm 0.008$ & $12.45  \pm 0.012$ & $12.344 \pm	0.017$  \\
            60477 & $13.21 \pm  0.008 $ & $12.536 \pm 0.038$ & $12.446 \pm 0.007$ & $12.335 \pm	0.011$  \\
            60481 & $13.313 \pm 0.029 $ & $12.523 \pm 0.011$ & $12.458 \pm 0.009$ & $12.341 \pm	0.014$  \\
            60484 & $13.272 \pm 0.013 $ & $12.575 \pm 0.011$ & $12.455 \pm 0.011$ & $12.341 \pm	0.014$  \\
            60486 & $13.347 \pm 0.006 $ & $12.591 \pm 0.005$ & $12.471 \pm 0.006$ & $12.356 \pm	0.01$  \\
            60489 & $13.384 \pm 0.005 $ & $12.63  \pm 0.004$ & $12.516 \pm 0.005$ & $12.404 \pm	0.009$  \\
            60493 & $13.428 \pm 0.009 $ & $12.63  \pm 0.011$ & $12.537 \pm 0.015$ & $12.336 \pm	0.021$  \\
            60496 & $13.475 \pm 0.008 $ & $12.66  \pm 0.007$ & $12.549 \pm 0.008$ & $12.405 \pm	0.017$  \\
            60517 & $14.312 \pm 0.01  $ & $13.178 \pm 0.007$ & $12.982 \pm 0.007$ & $12.774 \pm	0.013$  \\
            \hline
        \end{tabular}
    \end{table*}
\end{appendix}

\end{document}